\documentstyle[twocolumn,prl,aps,floats,graphicx]{revtex}

\begin{document}
\draft

\flushbottom
\title{Instability driven fragmentation of nanoscale fractal islands}

\author{
C. Br\'echignac$^{1}$, Ph. Cahuzac$^{1}$, F. Carlier$^{1}$, C. Colliex$^{1}$, 
J. Leroux$^{1}$, A. Masson$^{1}$, B. Yoon$^{2}$, U. Landman$^{2}$
}
\address{
$^{1}$Laboratoire Aim\'e Cotton, C.N.R.S. B\^at. 505, 
UPS, 91405 Orsay Cedex, France \\
$^{2}$School of Physics, Georgia Institute of Technology, Atlanta, 
Georgia 30332-0430 \\
}
\date{\today}
\maketitle
\begin{abstract}
Formation and evolution of fragmentation instabilities 
in fractal islands, obtained by deposition of silver clusters 
on graphite, are studied. 
The fragmentation dynamics and subsequent relaxation to the 
equilibrium shapes are controlled by the deposition conditions 
and cluster composition. 
Sharing common features with other materials' breakup phenomena, 
the fragmentation instability is governed by the length-to-width 
ratio of the fractal arms.
\end{abstract}
\pacs{PACS numbers: 68.55.Jk, 61.43.Hv, 81.16.Rf, 68.55.-a}

\narrowtext

The formation, evolution and consequences of instabilities in 
materials systems have been the subject of fruitful inquiry for a 
long time\cite{rayleigh,chandrasekhar,eggersRMP,mullins}. 
These investigations have been performed in various disciplines 
(e.g. hydrodynamics, solid state physics and materials science) and 
have been applied to systems characterized by different 
length-scales, from the macroscopic
\cite{rayleigh,chandrasekhar,eggersRMP,mullins} to the nanoscale
\cite{moseler} regimes. 
Such studies provide deep insights into fundamental issues 
pertaining to the interplay and balance governing the bulk and 
surface contributions to the energetics of materials, their 
morphological stability, and the relaxation dynamics in 
non-equilibrium complex systems. Furthermore, studies of such 
materials instabilities have important technological implications 
concerning the formation and control of materials systems of desired 
properties (e.g. nanojets\cite{moseler}, nanowires\cite{serena}, 
thin films and other surface supported structures).

Here we report on experimental and theoretical investigations of the 
development and evolution of morphological instabilities in supported 
fractal films, generated through the deposition of a narrow 
size-distribution of silver clusters on a graphite substrate at 
room temperature\cite{yoon,goldby}. 
We find that such deposited structures fragment (at ambient 
temperature) through a surface self-diffusion mechanism [4b], and 
consequently the mass density and its fractal dimension are conserved 
throughout the process. We demonstrate that the fragmentation 
dynamics and the relaxation to the equilibrium shape, can be 
controlled by the deposition conditions as well as by the composition 
of the incident clusters. 
Analysis of the fully fragmented fractal islands leads to 
identification of the ratio between the length and the width of the 
fractal branch (prior to fragmentation) as the critical parameter 
characterizing the instability that underlies the fragmentation. 
The emergence of such a characteristic parameter as well as its 
critical value (typically of the order of 4.5), implies that the 
fragmentation of surface-supported solid fractal structures studied 
here shares the same universality class as other instability driven 
fragmentation phenomena
\cite{rayleigh,chandrasekhar,eggersRMP,mullins}. 

Under appropriate conditions, films grown through the deposition of 
preformed clusters exhibit a non-equilibrium fractal pattern
\cite{yoon,jensen}. 
The fractal shape of the supported islands is a consequence of the 
diffusion limited aggregation (DLA) mechanism\cite{witten}, and 
studies pertaining to the morphological stability of the films 
provide valuable information about the intrinsic properties of these 
non-equilibrium objects. 
However, experimental reports about fractal stability are scarce
\cite{hwang,roder}. 
In these experiments, fractal islands relax upon thermal annealing 
into more compact forms and no fragmentation has been observed. 
Nevertheless, numerical simulations predicted a late (post-growth)
fragmentation stage, with periphery diffusion as the dominant 
mass transport mechanism\cite{irisawa,thouy,sempere,conti}. 
Our experiments provide evidence for fractal fragmentation within 
a time-scale that depends strongly on the fractal branch width and 
on the surface self-diffusion coefficient. 
The surface diffusion current is proportional to the local gradient 
of the fractal branch curvature
(\cite{mullins,thouy,eggers}, and Eq. (\ref{eq1}) below). 

Gas-phase neutral silver clusters, produced by a gas-aggregation 
cluster source\cite{yoon}, are deposited at low impact energy 
(0.05 eV/atom), on a room temperature cleaved graphite surface 
maintained at a high vacuum (10$^{-9}$ torr). 
The neutral cluster size-distribution is measured by a 
time-of-flight mass spectrometer. 
The distribution is characterized by a Gaussian with a peak at
150 atoms (i.e. clusters with a mean diameter of 2 nm) and a 
full width of half-maximum corresponding typically to 20 \% 
of the mean diameter value.
A crystal quartz micro-balance measures the flux of the incident 
neutral clusters (of the order of 10$^{10}$ clusters/cm$^2$s, with 
deposition times between 1 to 30 minutes). 
Since the per-atom kinetic energy of the incident clusters is very 
low compared with their binding energy (1.2 eV/atom) the clusters 
migrate on the surface as a whole and grow into islands\cite{yoon} 
which can continue to evolve past the formation stage. 
The prepared samples were either transferred in air and imaged in a 
scanning electron microscope (SEM), or transferred into high vacuum 
and imaged with a non-contact Atomic Force Microscope (AFM). 
The similarity between the images obtained in both cases, shows that 
the transfer in air does not affect the island morphologies. 
We verified by X-ray photoemission spectroscopy that further 
evolution of the deposited islands after transfer in air is hindered 
by surface oxidation. 
Therefore, we can define an observation time window (OTW) that starts 
from the beginning of the deposition and ends when the sample is 
exposed to air (typical observation time is of the order of one hour).

In previous studies we have shown that the morphologies of islands 
generated by cluster deposition are the consequence of the interplay 
between the coalescence time (that is, the time for an added cluster 
to merge into the island) and the aggregation time interval, $t_a$, 
which is the time-lapse between two successive arrivals of clusters 
to the island\cite{yoon}. 
When the coalescence time exceeds $t_a$ a ramified fractal shape 
develops with a constant branch width $l_0$. 
This value depends on the size of the incident clusters, on the flux 
and, to some extent, on the deposition time at high coverage. 
$l_0$ decreases for larger clusters and/or higher flux, and it 
increases with time at high coverage\cite{yoon}. 

%***************** begin figure 1 **************************
\begin{figure}[t]
\centering\includegraphics[width=7.5cm]{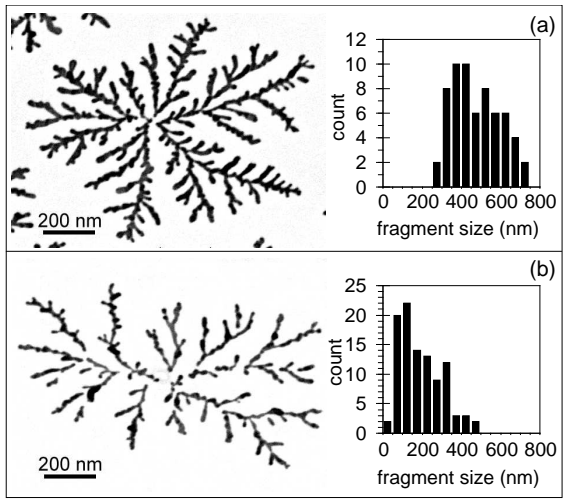}\\
~~~\\
\caption{
SEM images of silver fractal islands and histograms of their 
branch-lengths (on the right), obtained from deposition of 
Ag$_{150}$ clusters on graphite, for two different coverages: 
(a) 6 ML, and (b) 2 ML. 
\label{fig1}
}
\end{figure}
%***************** end figure 1 **************************
 
In the first set of experiments we illustrate the effect of the 
fractal branch-width on the fragmentation propensity. 
Figs.\ref{fig1}a and \ref{fig1}b show SEM images of typical 
individual Ag fractal islands (with the fractal dimension
\cite{witten} $D_f=1.7$).
The two coverages that we show were obtained by carrying out the 
deposition for 7 minutes, and subsequently blocking half of the 
substrate by a mask and continuing the deposition onto the unmasked 
half for an additional interval of 20 minutes. 
The islands shown in Figs \ref{fig1}a and \ref{fig1}b have 
approximately the same global size ($L \sim 1$ $\mu$m). 
At large coverage (6ML, unmasked part), the grown islands are 
characterized by thick branches ($l_0=25$ nm, see Fig. \ref{fig1}a) 
and they are not fragmented, except occasionally at the center of 
the island. 
On the other hand, at small coverage (2ML, masked part) the fractal 
islands show thinner branches ($l_0=15$ nm, see Fig. \ref{fig1}(b)) 
and, within our OTW, they  exhibit fragmentation throughout the 
island. 
Treating the fractal as a set of consecutive bars and focusing on 
the main branches (defined as the longest continuous quasi-linear 
segments), the histograms on the side of the images in 
Fig. \ref{fig1} depict the number distributions of bars as a 
function of the bar length, showing a shift to smaller length 
as the branch width decreases. 
For the system shown in Fig. \ref{fig1}b the fragments have not 
reached their equilibrium spherical shape. 
The non-observability of fragmentation in Fig. \ref{fig1}a and the 
incomplete fragmentation in Fig. \ref{fig1}b, are consequences of 
our OTW value and the slow room-temperature fragmentation kinetics 
(see below).

%***************** begin figure 2 **************************
\begin{figure}[t]
\centering\includegraphics[width=7.5cm]{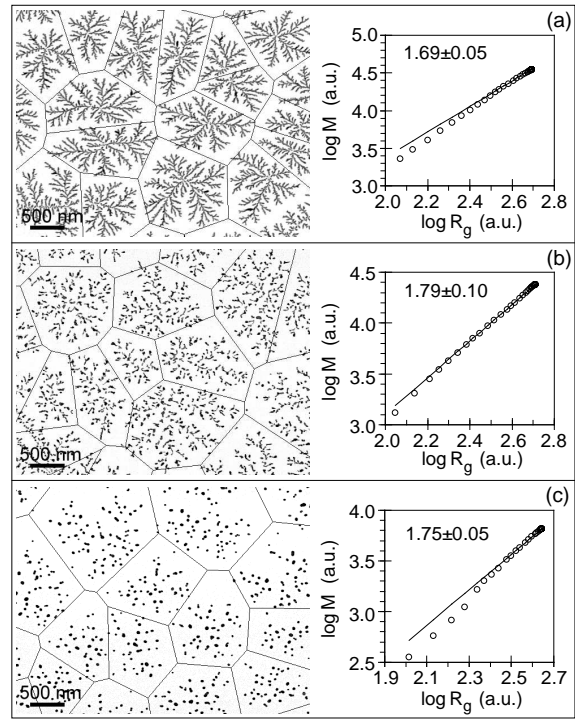}\\
~~~\\
\caption{
Influence of an impurity in the incident clusters on the stability 
of the fractal islands. 
The SEM image in (a) corresponds to deposition of Ag$_{100}$ 
clusters and those in 
(b) and (c) correspond to deposition of such clusters but with 0.5\% 
and 1\% oxygen impurity, respectively. 
In all the images a Voronoi polygon construction is superimposed. 
On the right of the images analysis[19] of the fractal 
dimension is displayed, and the asymptotic values are given.
\label{fig2}
}
\end{figure}
%***************** end figure 2 **************************

In a second set of experiments we vary the composition of the 
incident clusters by adding trace amounts of oxygen and 
water during the silver cluster formation stage (i.e. before 
deposition). 
Fig. \ref{fig2} shows SEM images of (multi-island) fractal films 
obtained, for the same coverage, through deposition of pure silver 
clusters (Fig. \ref{fig2}a), and via deposition of clusters with 
varying trace amount of oxygen (Figs. \ref{fig2}b and \ref{fig2}c). 
An overall characterization of the deposited film can be obtained 
through a 2D Voronoi polygon construction\cite{mulheran}, with each 
Voronoi cell being associated with a single island. 
The density of Voronoi polygons is determined by the nucleation 
sites (most likely defects) on the graphite surface 
(10$^8$ sites/cm$^2$). 
As aforementioned, islands obtained from deposition of pure clusters 
under the above conditions do not undergo fragmentation 
(Fig. \ref{fig2}a), whereas those resulting from clusters with 
variable trace amount of oxygen are fragmented. 
The measured density of fragments is the same for the data shown 
in Fig. \ref{fig2}b and \ref{fig2}c, and in the latter the shape 
relaxation of the fragments has been completed (resulting in 
spherical shape). 
Remarkably, rather similar asymptotic values of the mass fractal 
dimension\cite{hwang,fractaldim} ($D_f=1.70\pm0.10$, see right 
panels in Figs. \ref{fig2}(a-c)) were determined for both the pure 
and impurity-containing cluster depositions.

From the above observation of a common fractal dimension, the 
growth of a continuous fractal island should precede its subsequent 
fragmentation into many pieces. 
The fragmentation process was explored through examination of the 
correlation function between the fragments belonging to a given 
island. 
Such an analysis (performed for numerous islands) reveals a nearly 
constant nearest-neighbor distance $\lambda$ between fragments 
belonging to the same fractal branch. 
The histogram of the nearest-neighbor distance peaks at 
$\lambda=75\pm10$ nm. 
From the size distribution of the fragments we deduced a mean 
fragment diameter $2R=30\pm3$ nm. 
Treating the fractal as a set of uniform cylindrical branches, 
the width $l_0$ of a branch that would have existed prior to 
fragmentation can be estimated from the conservation of mass, 
i.e. $16R^3=3\lambda l_0^2$. 
The above considerations lead to a ratio $\lambda/l_0 = 4.8$. 
This value is in good agreement with that obtained from linear 
stability analysis, yielding an optimal wavelength of $4.443l_0$
[4c] for fragmentation of a solid cylinder induced by a capillary 
instability (through a surface self-diffusion mechanism), and 
interestingly it is close to the value predicted by Rayleigh for 
breakup of a cylindrical column of an inviscid liquid
\cite{rayleigh}.

To further investigate the growth and the breakup mechanism we 
performed simulations\cite{numethod} of the (axial) growth of a 
material with a 3D cylindrical symmetry, by adding spherical 
clusters to each side of a cylindrical structure (lying along the 
$z$ axis) at an aggregation time interval $t_a$. 
The evolution of the shape of the axisymmetric material structure
[4b] is described by
\begin{equation}
\frac{\partial h}{\partial t} = 
\frac{B}{\rho}\frac{\partial}{\partial s}
{\left (\rho\frac{\partial\kappa}{\partial s}\right )}
\label{eq1}
\end{equation}.
Here $\partial h/\partial t$ and $s$ are, respectively, the local 
transversal displacement per unit time and the curvilinear 
coordinate of the contour of the material structure on the 
$\rho-z$ plane of the cylindrical coordinate system ($\rho$ is the 
local radius of the cylinder along the $z$ axis). 
$\kappa$ is the local curvature expressed as 
$\kappa=\kappa_1+\kappa_2$, where $\kappa_1$ and $\kappa_2$ are the 
curvatures defined in the $\rho-z$ plane and in the plane normal to
that which is tangent to the surface of the material structure, 
respectively.  
The fourth-order diffusion constant $B$ \cite{mullins} is given by 
$B=D_s\gamma\Omega^2S_0/k_BT$ where $D_s$ is the surface 
self-diffusion coefficient, $\gamma$ is the surface tension, 
$\Omega$ is the atomic volume, and $S_0$ the number of atoms per 
unit surface area. 
The length and time units (tu) are selected by setting both the 
diameter of the incident spherical clusters, and the value of $B$, 
to unity.

For a constant aggregation time interval $t_a$ an elongated cylinder 
of nearly constant width, with semispherical caps at both ends, 
develops at the early growth stage. 
When the length of the structure exceeds a critical value an 
instability develops, resulting in fragmentation into a string of 
equidistant spheres (in agreement with the 2D simulations in ref. 
\cite{thouy}).

%***************** begin figure 3 **************************
\begin{figure}[t]
\centering\includegraphics[width=7.5cm]{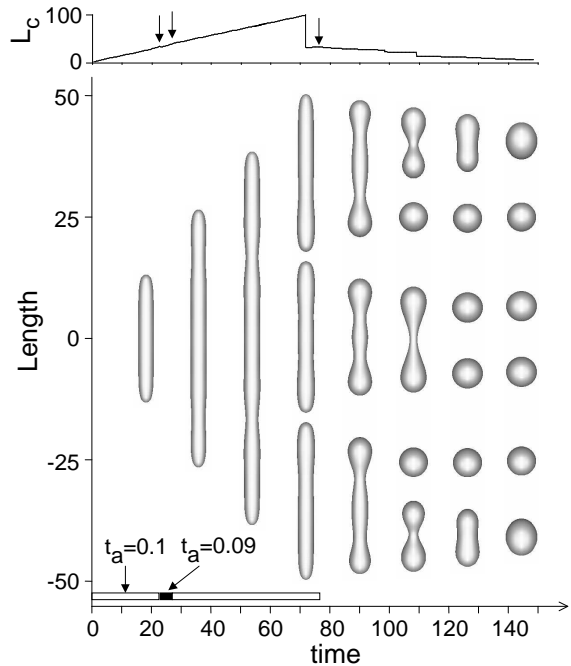}\\
~~~\\
\caption{
Results from 3D simulations, whose procedure is described in the 
text. 
In the bottom panel the shape of the evolving cylindrical structure 
and fragmentation events are depicted versus time. 
In the upper panel the length of the longest continuous part of the 
cylindrical structure (i.e. the intact length) is plotted versus the 
simulation time; each fragmentation event is signaled by the 
appearance of a saw-tooth shaped discontinuity, and arrows indicate 
a change in the clusters deposition flux. 
The addition of clusters is stopped for $t>78$ tu (rightmost arrow) 
The length and time units are scaled as described in the text.
\label{fig3}
}
\end{figure}
%***************** end figure 3 **************************

To model low-frequency experimental fluctuations of the cluster 
source, the flux of the incident clusters is varied in the 
simulations during growth. 
Typical results of such a simulation are shown in Fig. \ref{fig3}. 
In this particular simulation we used an aggregation time 
$t_a = 0.1$ tu for an initial period of 22 tu, followed by a 
period of 5 tu during which $t_a$ was reduced to a value of 0.09 tu, 
and subsequently $t_a$ was restored to its original value. 
At the early stage ($t<20$ tu), an elongated capped 
cylindrical shape of nearly constant width develops. 
As a result of the above flux fluctuation the cylinder develops 
a local perturbation of its width which culminates in fragmentation 
at $t=72$ tu. 
At $t= 76$ tu we stopped the addition of clusters. 
Subsequent evolution involves the development of morphological 
instabilities that lead to complete fragmentation into a string of 
spheres, which are not rigorously equidistant. 
It is of interest to remark on the evolution of the three fragments 
for $t>90$ tu. 
We observe that while eventually all three underwent binary 
fragmentation, that of the middle one is mass-symmentric and those 
of the two end ones are asymmetric. 
In fact, the morphological instability of the two end-most pieces 
(see the peanut shapes at $t=110$ tu) did not achieve the critical 
length-to-width ratio and the process culminated in relaxation into 
a larger-in-size spherical shape. 
Repeating the simulation for several arbitrary temporal fluctuations, 
one obtains a distribution of distances $\lambda$ between neighboring 
spherical fragments, peaking at a value of 4.5 $l_0$, where $l_0$ is 
the mean diameter of the cylindrical structure (prior to 
fragmentation), in agreement with the experimental histograms. 

Underlying the evolution of a morphological instability of an 
elongated solid 3D material structure is the dependence of the 
surface diffusion rate on the local curvature. 
A linear stability analysis [4c] of Eq. (\ref{eq1}), limited to a 
small axisymmetric disturbance (with wave-vector $k$) of the 
cylinder's radius, gives for the fastest growing disturbance 
$k^2\rho_0^2=0.5$, where $\rho_0$ is the radius of the unperturbed 
uniform cylinder; i.e., an optimal instability wavelength 
$\lambda_c= 2^{1/2}2\pi\rho_0\simeq4.443l_0$.
This value of $\lambda_c/l_0$ is very close to the aforementioned 
simulated value of $4.5l_0$. 
We remark that within linear stability analysis, such instability 
does not develop for a 2D system. 

From the measured mean radius of the relaxed fractal fragments 
(Fig. \ref{fig2}c), together with mass conservation, we conclude 
that both the fragmented and unfragmented fractal islands are 3D in 
nature, with a wetting angle in the vicinity of 
$\pi/2$\cite{mccallum}. 
The disturbance which may culminate in fragmentation can be either 
of thermal origin or caused by substrate induced stresses, and the 
relevant time-scale for fragmentation involving surface 
self-diffusion is $\tau=l_0^4/B$. 
This implies that the kinetics of fragmentation is influenced by
$l_0$ and/or $B$. 
Fig. \ref{fig1} demonstrates the influence of $l_0$. 
Using parameters for pure silver\cite{stoldt,akamatsu} we estimate 
at 300K, $\tau(l_0=25$nm$) = 2300$ s and $\tau(l_0=15$nm$) = 300$ s.
Consequently for a typical 2000 s OTW used by us, 
observation of room-temperature fragmentation is (kinetically) 
unlikely for the thicker branches (Fig. \ref{fig1}a), while it 
should be readily observed (at least partial fragmentation) for the 
thinner ones (Fig. \ref{fig1}b). Figs. \ref{fig2}b and \ref{fig2}c 
depict complete fragmentation within the same OTW, showing evidence 
for a faster fragmentation kinetics (i.e. decrease of $\tau$) with 
increasing trace amounts of oxygen in the incident clusters.

In summary, we found via experiments and theoretical analysis, that 
the room-temperature fragmentation instability of fractal islands 
formed through the deposition of silver clusters on graphite, is 
governed by the ratio of the length to the width of the fractal arms; 
such length-to-width control parameter is a common feature in many 
instability phenomena
\cite{rayleigh,chandrasekhar,eggersRMP,mullins,moseler}. 
In particular, the wavelength of the fastest growing instability is 
determined by this critical ratio (having a value of $\sim 4.5$), 
and the actual observability of fragmentation depends on the 
fragmentation kinetic, involving surface self-diffusion, and thus on 
the experimental OTW. 
The fragmentation dynamics and the relaxation to the equilibrium 
shapes can be controlled by the deposition conditions (including 
size selection of the incident clusters) as well as by the 
composition of the incident clusters. 
Technological implications may include issues pertaining to the 
growth of surface-supported elongated structures (nanowires), 
and the self-organization of structures made of supported 
nanoparticles with well-defined sizes and interparticle spacing.

This research is supported by CNRS and the US DOE 
(Grant No. FG05-86ER-45234). Computations were performed at 
the GIT Center for Computational Materials Science.


\begin{references}
\bibitem{fractaldim}The asymptotic mass fractal dimension was 
measured for large $R_g$ by using the power law relation between the 
mass $M$ and the gyration radius $R_g$.
\bibitem{numethod}The numerical method that we used is a 3D extension 
of the 2D simulation described in ref. \cite{thouy}. 
Moreover, in ref. \cite{thouy} only fragmentation after growth 
was simulated.
\end{references}
\end{document}